\documentclass[a4paper,english,12pt,amssymb,nofootinbib,superscriptaddress]{revtex4-2}
\usepackage{tcolorbox}
\usepackage{lmodern}
\usepackage{lmodern}

\usepackage[T1]{fontenc}
\usepackage[latin9]{inputenc}
\usepackage{babel}
\usepackage{mathrsfs}
\usepackage{amsmath}
\usepackage{amssymb}
\usepackage{esint}
\usepackage[unicode=true,pdfusetitle,
bookmarks=true,bookmarksnumbered=false,bookmarksopen=false,
breaklinks=false,pdfborder={0 0 1},backref=false,colorlinks=false]
{hyperref}

\makeatletter

\@ifundefined{textcolor}{}
{%
	\definecolor{BLACK}{gray}{0}
	\definecolor{WHITE}{gray}{1}
	\definecolor{RED}{rgb}{1,0,0}
	\definecolor{GREEN}{rgb}{0,1,0}
	\definecolor{BLUE}{rgb}{0,0,1}
	\definecolor{CYAN}{cmyk}{1,0,0,0}
	\definecolor{MAGENTA}{cmyk}{0,1,0,0}
	\definecolor{YELLOW}{cmyk}{0,0,1,0}
}

\usepackage{babel}
\usepackage{babel}
\usepackage{babel}
\usepackage{babel}
\usepackage{babel}
\usepackage{babel}
\usepackage{babel}
\usepackage{babel}
\usepackage{babel}
\usepackage{babel}
\usepackage{babel}
\usepackage{babel}
\usepackage{babel}
\usepackage{babel}
\usepackage{babel}
\usepackage{babel}
\usepackage{babel}
\usepackage{graphicx}

\def\be{\begin{equation}}
\def\ee{\end{equation}}

\@ifundefined{textcolor}{}{%
	\definecolor{BLACK}{gray}{0}
	\definecolor{WHITE}{gray}{1}
	\definecolor{RED}{rgb}{1,0,0}
	\definecolor{GREEN}{rgb}{0,1,0}
	\definecolor{BLUE}{rgb}{0,0,1}
	\definecolor{CYAN}{cmyk}{1,0,0,0}
	\definecolor{MAGENTA}{cmyk}{0,1,0,0}
	\definecolor{YELLOW}{cmyk}{0,0,1,0}
}

\usepackage{latexsym}\usepackage{bm}

\makeatother

\begin{document}

\title{A Systematic Construction of Kastor-Traschen Currents and their Extensions to Generic Powers of Curvature} 

\author{Zeynep Tugce Ozkarsligil}
\email{ozkarsligil.zeynep@metu.edu.tr}
\affiliation{Department of Physics,\\
 Middle East Technical University, 06800, Ankara, Turkey}

\author{Bayram Tekin}
\email{btekin@metu.edu.tr}

\affiliation{Department of Physics,\\
 Middle East Technical University, 06800, Ankara, Turkey}
\date{\today}

\begin{abstract} 
\noindent Kastor and Traschen constructed totally anti-symmetric conserved currents that are linear in the Riemann curvature in spacetimes admitting Killing-Yano tensors. The construction does not refer to any field equations and is built on the algebraic and differential symmetries of the Riemann tensor as well as on the Killing-Yano equation. Here we give a systematic generalization of their work and find divergence-free currents that are built from the powers of the curvature tensor. 
 A rank-4 divergence-free tensor that is constructed from the powers of the curvature tensor plays a major role here and it comes from the Lanczos-Lovelock theory. 
\end{abstract}

\maketitle


\section{\label{intro} Introduction}

In \cite{kt1} Kastor and Traschen (KT) introduced the following anti-symmetric tensor
\be
{\mathcal{J}}_{\text{KT}}^{\mu \nu} := -\frac{1}{4}\Big (f_{\sigma \rho} \, R^{\mu \nu \sigma \rho} - 2 f^{\mu}\,_{\sigma} \,
R^{\sigma \nu} + 2 f^{\nu}\,_{\sigma} \, R^{\sigma \mu} + f^{\mu \nu} \, R \Big),  \label{KT3}
\ee
where  $R^{\mu \nu\sigma \rho}$, $R^{\mu \nu}$, and $R$ are the Riemann, Ricci tensors, and the scalar curvature,  respectively.  Here  $f^{\mu \nu}$ is a Killing-Yano tensor; i.e. an anti-symmetric tensor satisfying the Killing-Yano equation 
\be
{\nabla}_{\mu} \, f_{\nu \sigma} 
+ {\nabla}_{\nu} \, f_{\mu \sigma} = 0 \, . \label{yaneq}
\ee
The interesting fact about ${\mathcal{J}}^{\mu \nu}$ is that it is a theory-independent conserved ``current''; i.e.
\begin{equation}
\nabla_\mu {\mathcal{J}}_{\text{KT}}^{\mu \nu}=0 
\label{conserved}
\end{equation}
for all smooth metrics satisfying the Bianchi Identity for the Riemann tensor and its contractions. So Einstein equations or any other field equations have not been used to show the covariant conservation of the current, hence (\ref{KT3}) is a geometric object on a generic manifold of dimension $D$ larger than 3. 
In lower dimensions, for $D=2$ and $D=3$, ${\mathcal{J}}_{\text{KT}}^{\mu \nu}$ vanishes identically.  Note that one can add a term to this current without destroying its properties, that term would be of the form $\alpha f^{\mu \nu}$, since $\nabla_\mu  f^{\mu \nu}=0$ with $\alpha$ an arbitrary constant or it can be chosen to be the cosmological constant $ \Lambda$ to comply with the linearity of the KT current in the curvature. We shall comment on this below.

Any conserved current on a manifold is both a curiosity and a valuable asset in constructing conserved quantities. In fact, in \cite {kt1}, generalizing the Arnowitt-Deser-Missner \cite{adm} and the Abbott-Deser \cite{ad} (Killing) charge constructions, (\ref{KT3}) was used to define conserved {\it mass density} for asymptotically flat spacetimes only in certain spatial directions which is the case, for example for $D$ or $p$ branes.  The total mass of these infinitely extended objects would be infinite, but their mass density is preserved and positive under certain conditions as was shown in \cite{kt2}. 
The Kastor-Traschen construction was extended to asymptotically transverse anti-de-Sitter spacetimes in \cite{Cebeci} following the formalism of \cite{dt1,dt2} and the first order formulation of \cite{Cebeci2}. [Note that one could naively worry that a conserved mass density would not allow any type of motion, that is not the case here. The mass density is calculated at infinity in certain transverse directions to the extended object; it does allow the motion of the object. For an explicit example, see the long Weyl rod computation in \cite{Cebeci}.]

The usefulness of a KT-type conserved current,  ${\mathcal{J}}_{\text{KT}}^{\mu \nu}$,\footnote{One should be happy any time one sees a covariantly conserved anti-symmetric object as they lead to conserved quantities; somebody said it before us: ``God loves anti-symmetry more than symmetry''.} is clear, but its construction and generalizations to higher powers in the curvature are not clear, because the way (\ref{KT3}) appeared in the original paper seems more like serendipity, and to be able to go beyond the linearity in the curvature and include more powers of curvature would be rather difficult without a systematic approach.  Here we shall remedy this and first show how (\ref{KT3}) appears rather naturally and how it can be extended to generic powers of curvature. 
So our goal here is the following: without referring to any field equations of a particular gravity theory, construct conserved anti-symmetric currents that are linear in the Killing-Yano tensor, but non-linear in the Riemann tensor and its contractions for smooth manifolds of dimensions $D \ge 4$.
For a recent nice summary of the uses of Killing-Yano tensors see \cite{sar}. 

In section II, we give a concise form of the KT current that is amenable to generalization; in section III we use the Lanczos-Lovelock theory to build a conserved rank-4 tensor that has the same algebraic properties of the Riemann tensor but does not obey the differential Bianchi Identity, yet it is divergence-free. In the Appendix, we give a differential form equivalent version of the discussion. 

\section{ The ${\mathcal{P}}$-tensor and the Kastor-Traschen Currents}

In \cite{Altas1, Altas2}, for the intent of writing the conserved charges in asymptotically AdS spacetimes in terms of the Riemann tensor, the authors introduced a $(1,3)$ rank tensor, the  $\text{\ensuremath{{\cal {P}}}}$-tensor which reads in generic $D \ge 4$ dimensions as follows:
\begin{equation}
\text{\ensuremath{{\cal 
{P}}}}^{\nu}\thinspace_{\mu\beta\sigma}:=R^{\nu}\thinspace_{\mu\beta\sigma}+\delta_{\sigma}^{\nu}\text{\ensuremath{{\cal {G}}}}_{\beta\mu}-\delta_{\beta}^{\nu}\text{\ensuremath{{\cal {G}}}}_{\sigma\mu}+\text{\ensuremath{{\cal {G}}}}_{\sigma}^{\nu}g_{\beta\mu}-\text{\ensuremath{{\cal 
{G}}}}_{\beta}^{\nu}g_{\sigma\mu}+\frac{R}{2}\left(\delta_{\sigma}^{\nu}g_{\beta\mu}-\delta_{\beta}^{\nu}g_{\sigma\mu}\right),
\label{P1}
\end{equation}
where  the (cosmological) Einstein tensor is defined as ${\cal {G}}_{\mu\nu}:= R_{\mu \nu} -\frac{1}{2}g_{\mu \nu} R+ \Lambda g_{\mu \nu}$. In what follows we will set $\Lambda =0$, but keeping it would not drastically alter the picture. The  $\text{\ensuremath{{\cal {P}}}}$-tensor has the following properties each of which can be easily checked from  its definition and the symmetries of the Riemann tensor and the Bianchi identities:
\begin{enumerate}

\item It vanishes identically in two and three dimensions.

\item It has the algebraic symmetries of the Riemann tensor, and satisfies the algebraic Bianchi Identity $\text{\ensuremath{{\cal {P}}}}_{\mu [\nu\beta\sigma]}=0$.

\item  Its trace yields, not the Ricci tensor, but the Einstein tensor
\begin{eqnarray}
\text{\ensuremath{{\cal 
{P}}}}^{\nu}\thinspace_{\mu\nu\sigma}= (3-D)\mathcal{G}_{\mu \sigma}.
\end{eqnarray}
\item It does not obey the differential Bianchi Identity, namely $\nabla_{[\mu} \text{\ensuremath{{\cal {P}}}}_{\rho \nu]\beta\sigma} \ne 0$, but it obeys the following covariant divergence-free property (for all of its indices)
\begin{equation}
\nabla_\nu \text{\ensuremath{{\cal {P}}}}^{\nu}\thinspace_{\mu\beta\sigma}=0.
\label{cov}
\end{equation}
Needless to say, the Riemann tensor does not have this property for generic spacetimes.
\item In four dimensions, it is equal to the {\it double dual } of the Riemann tensor:
\begin{equation}
\text{\ensuremath{{\cal {P}}}}^{\mu \nu\alpha\beta} = \star R\star^{ \mu \nu \alpha \beta} := \frac{1}{4}\epsilon^{\mu \nu \sigma \rho}  \epsilon^{\alpha \beta \delta \lambda} R_{ \sigma \rho \delta \lambda}, \qquad D=4.
\end{equation}
\item Its contraction with the Riemann tensor yields the Gauss-Bonnet scalar
\begin{equation}
\text{\ensuremath{{\cal {P}}}}^{\mu \nu\alpha\beta} R_{\mu \nu\alpha\beta} =  R_{\mu \nu\alpha\beta}  R^{\mu \nu\alpha\beta} -4 R_{\mu \nu} R^{\mu \nu} + R^2.
\end{equation}
\end{enumerate}
In (\ref{P1}), one can add the constant term $-\frac{(D+1)\Lambda}{D-1}\left(\delta_{\sigma}^{\nu}g_{\beta\mu}-\delta_{\beta}^{\nu}g_{\sigma\mu}\right)$ without destroying any of the above properties and use the cosmological Einstein tensor. That would make $\text{\ensuremath{{\cal 
{P}}}}^{\nu}\thinspace_{\mu\beta\sigma}$  vanish for maximally symmetric spacetimes, and make it reduce to the Weyl tensor for all Einstein spacetimes.  
Here, we have not added that term. 

 Let us note that, in a rather surprising way, this tensor also appeared in a new definition of the surface gravity and the associated Hawking temperature of black holes \cite{Altas3}.

Let us now introduce another property of this tensor, which will help us prove the claims of this paper: the KT current (\ref{KT3}) is given as 

\begin{tcolorbox}
\begin{equation}
{\mathcal{J}}_{\text{KT}}^{\mu \nu} = -\frac{1}{4}\text{\ensuremath{{\cal {P}}}}^{\mu \nu}\,_{\sigma\rho} f^{\sigma \rho}.
\label{KTguzel}
\end{equation}
\end{tcolorbox}
\noindent This equation is easy to prove as one just uses the definition of the ${\mathcal{P}}$-tensor (\ref{P1}) and the anti-symmetry of the Killing-Yano tensor. 
Moreover, the covariant divergence property follows immediately, since
\begin{equation}
\nabla_\mu{\mathcal{J}}_{\text{KT}}^{\mu \nu} = -\frac{1}{4}\nabla_\mu \left(\text{\ensuremath{{\cal {P}}}}^{\mu \nu}\,_{\sigma\rho} \right) f^{\sigma \rho}-\frac{1}{4}\text{\ensuremath{{\cal {P}}}}^{\mu \nu}\,_{\sigma\rho} \nabla_\mu f^{\sigma \rho}.
\end{equation} 
The first term on the right-hand side vanishes due to (\ref{cov}), and the second term on the right-hand side vanishes since it can be written as  ${\mathcal{P}}^\nu\,_{[\mu 
\sigma \rho]} \nabla^\mu f^{\sigma \rho}$, due to the total anti-symmetry of the Killing-Yano tensor. This term is identically zero as the first one since ${\mathcal{P}}^\nu\,_{[\mu 
\sigma \rho]}=0$. Note that by adding a term proportional to  $\Lambda\left(\delta_{\sigma}^{\nu}g_{\beta\mu}-\delta_{\beta}^{\nu}g_{\sigma\mu}\right)$ in the ${\mathcal{P}}$-tensor, one can also generate the linear term $\alpha f^{\mu \nu}$ in the KT current that we discussed at the end of the paragraph that includes (\ref{conserved}).  

Once we have an anti-symmetric conserved current, it is easy to build total conserved charges on a manifold 
${\mathcal{M}}$ with a boundary as follows:  one defines a two-form in local coordinates
\begin{equation}
{\mathcal{J}}_{\text{KT}}:= \frac{1}{2} {\mathcal{J}}^{\text{KT}}_{\mu \nu} dx^\mu \wedge dx^\nu.
\end{equation}
This 2-form is not closed, but it yields a natural closed $D-2$ form, $\star {\mathcal{J}}_{\text{KT}}$, i.e. $d \star {\mathcal{J}}_{\text{KT}}=0$. Then the existence of a closed $D-2$ form yields the number  $Q:=\int_{\Sigma }\star {\mathcal{J}}_{\text{KT}}$ where $\Sigma$ is a co-dimension 2 submanifold of the spacetime, and the number corresponds to the de Rham period or the homology class of the submanifold \cite{kt1,Benn}.

One generalization of the rank-2 KT current was already done in the original work \cite{kt1} to rank-$n$ currents using a rank $n \le D$ Yano tensor $f_{\mu_1...\mu_n}=f_{[ \mu_1...\mu_n ]}$ that satisfies 
\be
{\nabla}_{\alpha} \, {f}_{\beta \mu_2 \dots \mu_n} 
+ {\nabla}_{\beta} \, {f}_{\alpha \mu_2 \dots \mu_n} = 0, \,  \label{nyaneq}
\ee
and yields the covariantly conserved $n$-current
\be
{\mathcal {J}}^{\mu_1 \dots \mu_n} = (n-1) \, R^{[\mu_1 \mu_2}\,_{\rho \sigma} \, f^{\mu_3 \dots \mu_n]\rho\sigma}
+ 4 (-1)^{n} \, R_{\sigma}\,^{[\mu_1} \, f^{\mu_2 \dots \mu_n]\sigma}
+ \frac{2}{n} \, R \, f^{\mu_1 \dots \mu_n},
\label{bigcurrent}
\ee
which can be written as \cite{kt1}
\begin{equation}
{\mathcal{J}}^{\mu_1 \dots \mu_n} = -N_n \delta_{\nu_{1}\dots\nu_{n}\sigma \rho }^{\mu_{1}\dots\mu_{n}\alpha \beta } f^{\nu_{1}\dots\nu_{n}} R^{\sigma \rho}_{\alpha \beta}, \label{cok}
\end{equation}
with $R^{\sigma \rho}_{\alpha \beta} \equiv R^{\sigma \rho}\,_{\alpha \beta}$, $N_n$ is a normalization constant, and the generalized Kronecker delta reads as
\begin{equation}
\delta_{\nu_{1}\dots\nu_{m}}^{\mu_{1}\dots\mu_{m}}=\det\left|\begin{array}{ccc}
\delta_{\nu_{1}}^{\mu_{1}} & \dots & \delta_{\nu_{1}}^{\mu_{m}}\\
\vdots & \ddots & \vdots\\
\delta_{\nu_{m}}^{\mu_{1}} & \dots & \delta_{\nu_{m}}^{\mu_{m}}
\end{array}\right|.\label{eq:gen_kronecker}
\end{equation}
For the case of rank-2 Yano tensor  (\ref{bigcurrent}), one has 
\begin{equation}
{\mathcal{J}}^{\mu_1\mu_2} = -\frac{1}{16} \delta_{\nu_{1}\nu_{2}\sigma \rho }^{\mu_{1}\mu_{2}\alpha \beta } f^{\nu_{1}\nu_2} R^{\sigma \rho}_{\alpha \beta},
\end{equation}
which is consistent with (\ref{KTguzel}) since one can also write the $\mathcal{P}$ tensor as 
\begin{equation}
\text{\ensuremath{{\cal {P}}}}^{\alpha\beta}_{\rho \sigma} =  \frac{1}{4}\delta^{\alpha \beta \lambda \gamma}_{\rho \sigma\mu \nu} R^{\mu \nu}_{ \lambda \gamma},\label{kisaP}
\end{equation}
as one can check by expanding the generalized Kronecker delta in terms of the determinant of Kronecker deltas. The above form of the current gave us a hint for generalizations that are non-linear in the curvature tensor which we perform in the next section. 

\section{Generalized Kastor-Traschen currents from Lanczos-Lovelock Theory }

\subsection{Construction of the generalized current}

In $D$ spacetime dimensions, the Lovelock gravity, or perhaps more properly Lanczos-Lovelock, gravity,  \cite{Lovelock1,Lovelock, Lanczos} is
defined by the Lagrangian \footnote{ We follow the notation of \cite{spectra} but multiply the Lagrangian by $\frac{1}{2^n}$. } 
\begin{equation}
\mathcal{L}_{\text{LL}}\left(R_{mn}^{kl}\right):=\sum_{n=0}^{\left[\frac{D}{2}\right]}a_{n}\mathcal{L}_{n},\label{eq:Lovelock}
\end{equation}
where $a_{n}$'s are dimensionful constants, and $\left[\frac{D}{2}\right]$
corresponds to the integer part of its argument. Each part is given as 
\begin{equation}
\mathcal{L}_{n}:=\frac{1}{2^n}\delta_{\nu_{1}\dots\nu_{2n}}^{\mu_{1}\dots\mu_{2n}}\prod_{p=1}^{n}R_{\mu_{2p-1}\mu_{2p}}^{\nu_{2p-1}\nu_{2p}},\label{eq:Lovelock_order}
\end{equation}
where the generalized Kronecker delta is defined as (\ref{eq:gen_kronecker}). By definition, the lowest order term is defined as the cosmological constant: $\mathcal{L}_{0}:= -2 \Lambda$. One can compute the next few terms: $n=1$ gives the Einstein-Hilbert, and $n=2$ gives the Gauss-Bonnet Lagrangians.  If the spacetime dimension $D$ is even, then the highest order term $\mathcal{L}_{D/2}$ is a pure divergence and does not contribute to the field equations; i.e. it is a topological invariant for compact manifolds. [See \cite{Gurses1, Gurses2} for a somewhat detailed discussion of the non-existence of the Einstein-Gauss-Bonnet theory in $D=4$ dimensions  that gathered so much recent attention.] The virtues of the Lanczos-Lovelock theory are well-known, and we shall not repeat them here, they are summarized in \cite{spectra,sah}. Our main intention here is to use this theory to first define a proper generalization of the ${\mathcal{P}}$-tensor of the previous section that has the same divergence-free property and the algebraic symmetries, but it is non-linear in the curvature tensor and its traces. If we can do that, it should be clear to the astute reader by now, we can generalize the KT currents. 

Therefore, from (\ref{eq:Lovelock_order}), let us define the generalized, still rank-4,  ${\mathcal{P}}$-tensor as\footnote{An equivalent tensor was constructed in \cite{kastorR} where it was called the Riemann-Lovelock curvature tensor.}
\begin{equation}
 {\mathcal{P}}_{(n)}^{\mu \nu \alpha \beta} R_{\mu \nu \alpha \beta}:=\frac{1}{2^n}\delta_{\nu_{1}\dots\nu_{2n}}^{\mu_{1}\dots\mu_{2n}}\prod_{p=1}^{n}R_{\mu_{2p-1}\mu_{2p}}^{\nu_{2p-1}\nu_{2p}}, \label{20denk}
\end{equation} 
or more explicitly,
\begin{tcolorbox}
\begin{equation}
 {\mathcal{P}}_{(n) \alpha \beta}^{\mu \nu}=\frac{1}{2^n}\delta_{\alpha \beta \nu_{3}\dots \nu_{2n}}^{\mu \nu\mu_{3}\dots \mu_{2n} }
 \prod_{p=2}^{n} R_{\mu_{2p-1}\mu_{2p}}^{\nu_{2p-1}\nu_{2p}}. \label{genP}
\end{equation} 
\end{tcolorbox}
Observe that  with the above normalization ${\mathcal{P}}_{(2) \alpha \beta}^{\mu \nu}$  corresponds to our earlier definition (\ref{P1}) or (\ref{kisaP}); and ${\mathcal{P}}_{(0) \alpha \beta}^{\mu \nu} = \frac{1}{2}\delta_{\alpha \beta}^{\mu \nu}$.  It is clear from the definition (\ref{genP}) that ${\mathcal{P}}_{(n) \alpha \beta}^{\mu \nu}$ satisfies the algebraic symmetries of the Riemann tensor and the first Bianchi Identity ${\mathcal{P}}_{(n) \alpha [\beta\mu \nu]}=0$. But we need to show that it is divergence-free. Let us work this out by direct computation:
\begin{eqnarray}
\nabla_\nu {\mathcal{P}}_{(n) \alpha \beta}^{\mu \nu} &=& \frac{n-1}{2^n} \delta_{\alpha \beta \nu_{1}\dots \nu_{2n}}^{\mu \nu\mu_{3}\dots \mu_{2n} }
\nabla_\nu \Big ( R^{\nu_{3} \nu_{4}}_{\mu_{3} \mu_{4}} \Big)
 \prod_{p=3}^{n} R_{\mu_{2p-1}\mu_{2p}}^{\nu_{2p-1}\nu_{2p}} \nonumber \\
 &=&  \frac{1}{3}\frac{n-1}{ 2^n } \delta_{\alpha \beta \nu_{1}\dots \nu_{2n}}^{\mu \nu\mu_{3}\mu_{4}\dots \mu_{2n} }
\nabla_{ [\nu} R^{\nu_{3} \nu_{4} }_{\mu_{3} \mu_{4}]}
 \prod_{p=3}^{n} R_{\mu_{2p-1}\mu_{2p}}^{\nu_{2p-1}\nu_{2p}} =0,
\end{eqnarray} 
where in the last line we used the differential Bianchi Identity on the Riemann tensor. So this construction generalizes all the features of ${\mathcal{P}}_{(2) \alpha \beta}^{\mu \nu}$ to ${\mathcal{P}}_{(n) \alpha \beta}^{\mu \nu}$ . Therefore, we can now define, conserved generalized KT rank-2 current as
\begin{tcolorbox}
\begin{equation}
{\mathcal{J}}_{(n)}^{\mu \nu} := -\frac{1}{4}\text{\ensuremath{{\cal {P}}}}^{\mu \nu}_{(n)\sigma\rho} f^{\sigma \rho} =-\frac{1}{4}\frac{1}{2^n}\delta_{\sigma \rho \nu_{3}\dots \nu_{2n}}^{\mu \nu\mu_{3}\dots \mu_{2n} } f^{\sigma \rho}
 \prod_{p=2}^{n} R_{\mu_{2p-1}\mu_{2p}}^{\nu_{2p-1}\nu_{2p}},
\label{KTsexy}
\end{equation}
\end{tcolorbox}
\noindent which is anti-symmetric and conserved $\nabla_\mu {\mathcal{J}}_{(n)}^{\mu \nu}=0$. Note that any linear combination of these currents in the form 
\begin{equation}
{\mathcal{J}}_{(\text{Total})}^{\mu \nu} :=\sum_{n=0}^{\left[\frac{D}{2}\right] }c_n {\mathcal{J}}_{(n)}^{\mu \nu}, \label{toplam}
\end{equation}
where $c_n$ are dimensionful constants.
\subsection{Wald's rank-4  tensor versus the $\mathcal{P}$ tensor}

Let us comment on the connection between the $\mathcal{P}$ tensor discussed above and a rank-4 tensor that appears in the Wald entropy computations in higher derivative gravity models \cite{Wald1, Wald2}. \footnote{We thank a conscientious referee for reminding us of this tensor.} Given a diffeomorphism invariant action built on the metric tensor and the Riemann tensor and its derivatives and contractions, $I= \int d^D x \sqrt{-g} \mathcal{L}(g_{\mu \nu}, R_{\mu \nu \alpha \beta}, \nabla_\sigma R_{\mu \nu \alpha \beta}, ...)$, the field equations suggest that one defines the following tensor that has the same {\it algebraic } symmetries as the Riemann tensor
\begin{equation}
 P^W_{\mu \nu \alpha \beta} := \frac{\partial \mathcal{L}}{\partial R^{\mu \nu \alpha \beta}},
\end{equation}
where we put a $W$ subscript that refers to Wald. Generically this tensor is {\it not} covariant divergence-free (i.e. $\nabla^\mu P^W_{\mu \nu \alpha \beta} \ne 0$), nor does it obey the differential symmetries of the Riemann tensor.  Recall that the covariant divergence-free property  (\ref{cov}) of our  $\mathcal{P}$ was a necessary ingredient in the construction of the above currents.  So generically  $P^W \ne \mathcal{P}$  and one cannot use Wald's tensor to extend the KT currents straightforwardly.  To demonstrate what we have just stated, let us consider a particular form of the action for which the Lagrangian is a polynomial in the Riemann tensor only, but does not depend on its derivatives. So we have  $I= \int d^D x \sqrt{-g} \mathcal{L}(g_{\mu \nu}, R_{\mu \nu \alpha \beta})$, of which the field equations are \cite{Tahsinfield} (here we use a more compact notation)
\begin{equation}
\mathcal{E}_{\mu \nu} = \nabla^\lambda \nabla^\sigma  P^W_{\mu \sigma \nu \lambda} + \nabla^\lambda \nabla^\sigma  P^W_{\nu \sigma \mu \lambda}+ \frac{1}{2} \left(  P^W_{\rho \sigma \lambda \nu} R^{\rho \sigma\lambda}\,_\mu + P^W_{\rho \sigma \lambda \mu} R^{\rho \sigma\lambda}\,_\nu \right )- \frac{1}{2} g_{\mu \nu} \mathcal{L}, \label{sk}
\end{equation}  
which is generically a fourth-order theory in terms of the dynamical field (the metric tensor). At this stage, one could try to restrict the set of theories by demanding that one should have a second-order theory just like Einstein's gravity, then this can be achieved by setting 
\begin{equation}
 \nabla^\sigma  P^W_{\mu \sigma \nu \lambda} = 0, \quad \text{we demand this}, 
\end{equation}
which reduces  (\ref{sk}) to a second order theory. Say we have no matter fields, then from  the reduced form of  (\ref{sk}), one obtains
 \begin{equation}
 \mathcal{L} = \frac{2}{D}  P^W_{\rho \sigma \lambda \mu} R^{\rho \sigma\lambda\mu}.
 \end{equation}
Finally, comparing with (\ref{20denk}), only in this case (which is the case of Lanczos-Lovelock theories), the Wald tensor and the $\mathcal{P}$ of this work are proportional to each other as
\begin{equation}
\mathcal{P}_{\mu \nu \alpha \beta} = \frac{2}{D} P^W_{\mu \nu \alpha \beta}. 
\end{equation}
Let us remark that  for this case, the Wald entropy of a black hole as a conserved charge of diffeomorphisms (computed in the spatial cross-section of a null horizon ) reads \cite{dewit}
\begin{equation}
S_W =  2 \pi \int_{S} P^W_{ \mu \nu \alpha \beta} \epsilon^{\mu \nu} \epsilon^{\alpha \beta} d^{D-2}S,
\end{equation}
where $ \epsilon^{\alpha \beta}$ are the binormal vectors to $S$ that involve the time-like Killing vector (see \cite{Ozen} for further details). For a physical interpretation of the conserved charges built in this work, this connection between the Wald entropy would be very valuable: but one has to be careful, in the former we use Killing-Yano tensors, while in the latter the Killing vectors.  

\subsection{Linearized KT currents}

As a separate note, to further gain insight into the physical meaning of the currents constructed here, we can consider {\it asymptotically} flat spacetimes that only have asymptotic (not exact) Killing-Yano tensors as was done in \cite{kt1}.  This vantage point as advocated in \cite{AD} in the case of Killing vectors leads to conserved charges (with respect to the background of which the charges are assumed to be zero), and it proceeds by the linearization of all relevant tensors about the background spacetime (with the metric $\bar{g}_{\mu \nu}$). The details of this construction can be found in the recent review \cite{rev}. Let $\bar{f}_{\mu \nu}$ be the background Killing-Yano tensor, then linearization of  (\ref{toplam}) around the background flat spacetime (with $\bar{R}_{\mu \nu \alpha \beta}=0$); and each term in (\ref{toplam}) is given as in (\ref{KTsexy}).  Then one realizes that the only contribution comes from the original Kastor-Traschen current that is linear in the curvature tensor, all other terms are built with at least two powers of the Riemann tensor and hence vanish for asymptotically flat backgrounds.  So the current construction here reduces to that of KT for all asymptotically flat spacetimes.  As an antisymmetric rank two current, the linearized version of the KT current in asymptotically flat spacetimes can be written in terms of potential as $ \left ({\mathcal{J}}_{(2)}^{\mu \nu} \right)_L = \bar{\nabla}_\sigma \ell^{\sigma \mu \nu}$, where the potential was given in {\cite{kt1}, and $\bar\nabla_\sigma$ denotes the covariant derivative with respect to the flat background.  On the other hand, if the background spacetime is not asymptotically flat, but say, asymptotically anti-de-Sitter, the terms in (\ref{toplam}) contribute to the total current depending on the number of dimensions $D$, since now  $\bar{R}_{\mu \nu \alpha \beta} \ne 0$. Let us show this with an example in AdS spacetime.
Linearizing (\ref{KTsexy}) about an AdS background yields
\begin{equation}
\left ({\mathcal{J}}_{(n)}^{\mu \nu} \right)_L = -\frac{1}{4}\left(\text{\ensuremath{{\cal {P}}}}^{\mu \nu}_{(n)\sigma\rho}\right)_L \bar{f}^{\sigma \rho}, 
\label{KTsexylin}
\end{equation}
where, from (\ref{genP}), we have 
\begin{eqnarray}
\left( {\mathcal{P}}_{(n) \alpha \beta}^{\mu \nu}\right)_L&=&\frac{(n-1)}{2^n}\delta_{\alpha \beta \nu_{3}\dots \nu_{2n}}^{\mu \nu\mu_{3}\dots \mu_{2n} }\left ({R}_{\mu_{2n-1}\mu_{2n}}^{\nu_{2n-1}\nu_{2n}} \right)_L
 \prod_{p=2}^{n-1} \bar{R}_{\mu_{2p-1}\mu_{2p}}^{\nu_{2p-1}\nu_{2p}}.
\end{eqnarray}
 For the background we have
\begin{equation}
\bar{R}_{\alpha \beta}^{\mu\nu}=\frac{2\Lambda}{\left(D-1\right)\left(D-2\right)}\left(\delta_{\alpha}^{\mu}\delta_{\beta}^{\nu}-\delta_{\beta}^{\mu}\delta_{\alpha}^{\nu}\right).
\end{equation}
For the sake of simplicity, let us consider the $n=3$ case (corresponding to the cubic Lanczos-Lovelock theory) for which we have to compute 
\begin{eqnarray}
\left( {\mathcal{P}}_{(3) \alpha \beta}^{\mu \nu}\right)_L&=&\frac{1}{2^2}\delta_{\alpha \beta \nu_{3}\dots \nu_{6}}^{\mu \nu\mu_{3}\dots \mu_{6} }\left ({R}_{\mu_{5}\mu_{6}}^{\nu_{5}\nu_{6}} \right)_L \bar{R}_{\mu_{3}\mu_{4}}^{\nu_{3}\nu_{4}},  
\end{eqnarray}
which reduces to
\begin{eqnarray}
\left( {\mathcal{P}}_{(3) \alpha \beta}^{\mu \nu}\right)_L&=&\frac{\Lambda}{(D-1)(D-2)}\delta_{\alpha \beta \mu_{3}\mu_{4} \nu_{5} \nu_{6}}^{\mu \nu\mu_{3}\mu_{4}\mu_{5} \mu_{6} }\left ({R}_{\mu_{5}\mu_{6}}^{\nu_{5}\nu_{6}} \right)_L.
\end{eqnarray}
We can reduce the contracted generalized Kronecker delta with six up and six down indices as follows  (see \cite{spectra})
\begin{equation}
\delta_{\alpha \beta \mu_{3}\mu_{4} \nu_{5} \nu_{6}}^{\mu \nu\mu_{3}\mu_{4}\mu_{5} \mu_{6}} =\frac{ (D-4)!}{ (D-6)!} \delta_{\alpha \beta  \nu_{5} \nu_{6}}^{\mu ,\nu\mu_{5} \mu_{6}}, \quad D \ge 6
\end{equation}
which then yields 
\begin{eqnarray}
\left( {\mathcal{P}}_{(3) \alpha \beta}^{\mu \nu}\right)_L&=&\frac{\Lambda}{(D-1)(D-2)}\frac{ (D-4)!}{ (D-6)!} \delta_{\alpha \beta  \nu_{5} \nu_{6}}^{\mu \nu\mu_{5} \mu_{6}}\left ({R}_{\mu_{5}\mu_{6}}^{\nu_{5}\nu_{6}} \right)_L.
\end{eqnarray}
The resulting computation is still a little bit tedious, one can show that
\begin{align}
\left ({\mathcal{J}}_{(3)}^{\mu \nu} \right)_L=\frac{4\Lambda}{(D-1)(D-2)}\frac{ (D-4)!}{ (D-6)!} \left ({\mathcal{J}}_{(2)}^{\mu \nu} \right)_L,
\end{align}
where  $\left ({\mathcal{J}}_{(2)}^{\mu \nu} \right)_L$ was shown in \cite{Cebeci} to be written in terms of a rank-3 potential as follows
\be
\left ({\mathcal{J}}_{(2)}^{\mu \nu} \right)_L =  3! \, {\bar{\nabla}}_{\sigma} \, 
\Big( \bar{f}^{\rho[\mu} \, {\bar{\nabla}}^{\nu} \, h^{\sigma]}\,_{\rho} 
+ \frac{1}{2} \, \bar{f}^{[\nu \mu} \, {\bar{\nabla}}^{\sigma]} \, h 
+ \frac{1}{2} \, h_{\rho}\,^{[\nu} \, {\bar{\nabla}}^{\sigma} 
\, \bar{f}^{|\rho|\mu]} 
- \frac{1}{2} \, \bar{f}^{[\nu \mu} \, {\bar{\nabla}}_{|\rho|} 
\, h^{\sigma]\rho}
+ \frac{1}{6} \, h \, {\bar{\nabla}}^{[\nu} \, 
\bar{f}^{\sigma\mu]} \Big) \, , \label{yan}
\ee
where the brackets denote antisymmetrization; and $h_{\mu \nu}:= g_{\mu \nu} - \bar{g}_{\mu \nu}$ is the deviation from AdS which is assumed to be small asymptotically at spatial infinity.  This form can easily be integrated to give conserved charges which we shall study more explicitly in a separate work. So the upshot is that each term in (\ref{toplam}) contributes to the KT current for asymptotically non-flat geometries, while only the original KT current contributes for asymptotically flat geometries akin to the case for the Killing charges which is just the Arnowitt-Deser-Misner \cite{adm} energy for asymptotically flat geometries even in higher curvature theories, but higher curvature terms contribute for asymptotically non-flat ones \cite{dt1,dt2}.
\section{Conclusions}

In this work, our goal was to generalize the Kastor-Traschen current {\cite{kt1}) that is linear in the Riemann curvature and the Killing Yano-Tensor to currents non-linear in the curvature. For this purpose, we reformulated the original KT current in terms of a rank-4 divergence-free tensor; and used that expression and the Lanczos-Lovelock theory to build the generalized current.  In the construction of these currents, the only ingredients are the Bianchi Identities of the Riemann tensor as well as the Killing-Yano tensor. The currents are valid for all spacetimes that admit Killing-Yano tensors. There are various possible extensions of our work: we used Killing-Yano tensors here, but one can extend the construction to the spacetimes admitting Conformal Killing-Yano tensors \cite{Linds}.

\appendix

\section{ The ${\mathcal{P}}$ tensor in the differential forms }
Here we briefly give the discussion of the  ${\mathcal{P}}$ tensor in terms of differential forms. For this purpose, we need to recast the Lanczos-Lovelock Lagrangians as $D$-forms first. 

The Cartan Structure equations for the curvature 2-form and the torsion 1-form read, respectively, as 
\begin{eqnarray} 
&&R ^a\,_b := d \omega ^a\,_b +\omega ^a\,_c \wedge \omega ^c\,_b, \\
&&T^a := d e ^a\,_b +\omega ^a\,_b \wedge e^b = D e^a.
\end{eqnarray}
The Bianchi identities are 
\begin{equation}
D \wedge R ^a\,_b =0, \hskip  1 cm  D \wedge T ^a - R ^a\,_b \wedge e^b =0,
\end{equation}
while the Lanczos-Lovelock Lagrangian as a function of the vierbein and the spin connection reads 
\begin{equation}
{\mathcal{L}}_{LL}[e^a, \omega^{ab}]  =  \sum_{n =0}^{[\frac{D}{2}]} a_n {\mathcal{L}}_{(n)},
\end{equation}
 where the $D$-form  Lagrangian for each $n$ is given as 
\begin{equation}
{\mathcal{L}}_{(n)} := \epsilon_{a_1...a_D} R^{a_1 a_2}\wedge...\wedge R^{a_{2 n-1} a_{2 n}}\wedge e^{a_{2 n+1}}\wedge..\wedge e^{a_{D}}.
\end{equation}
Then we can define the tensor-valued ${\mathcal{P}}$ $(D-2)$ form
\begin{equation}
 {\mathcal{P}}_{a_1a_2}:=   \epsilon_{a_1...a_D} R^{a_3 a_4}\wedge...\wedge R^{a_{2 n-1} a_{2 n}}\wedge e^{a_{2 n+1}}\wedge..\wedge e^{a_{D}}.
\end{equation}
We did not need them, but for completeness, let us give the field equations for the full theory:
The field equation coming from $ \frac {\delta I_{LL}[e^a, \omega^{ab}]}{ \delta e^b}$ reads
\begin{equation}
{\cal E}_b = \sum_{n =0}^{\left[\frac{D-1}{2}\right]} (D-2n)a_n {\cal E}_b^{(n)}  =0
\end{equation}
where 
\begin{equation}
 {\cal E}_b^{(n)} := \epsilon_{b a_2..a_{D-1}} R^{a_1 a_2}\wedge...\wedge R^{a_{2 n-1} a_{2 n}}\wedge e^{a_{2 n+1}}\wedge..\wedge e^{a_{D-1}}.
\end{equation}

The field equation coming from $ \frac {\delta I_{LL}[e^a, \omega^{ab}]}{ \delta \omega^{bc}}$ reads

\begin{equation}
{\cal H}_{b c} = \sum_{n =1}^{[\frac{D-1}{2}]} n(D-2n)a_n  {\cal H}_{b c}^{(n)} 
\end{equation}
where 

\begin{equation}
 {\cal H}_{b c}^{(n)} := \epsilon_{b c a_3..a_{D}} R^{a_3 a_4}\wedge...\wedge R^{a_{2 n-1} a_{2 n}}\wedge T^{a_{2 n+1}}\wedge..\wedge e^{a_{D-1}}
\end{equation}
and we observe that the torsion entered into the last equation which can be set to zero. See \cite{zan} for further details.

\end{document}